\RequirePackage[l2tabu,orthodox]{nag}

\documentclass[headsepline,footsepline,footinclude=false,oneside,fontsize=11pt,paper=a4,listof=totoc,bibliography=totoc,DIV=12]{scrbook} 

\PassOptionsToPackage{table,svgnames,dvipsnames}{xcolor}

\usepackage[utf8]{inputenc}
\usepackage[T1]{fontenc}
\usepackage[sc]{mathpazo}
\usepackage[ngerman,english]{babel} 
\usepackage[autostyle]{csquotes}
\usepackage[%
  backend=biber,
  url=true,
  style=numeric, 
  sorting=none, 
  maxnames=4,
  minnames=3,
  maxbibnames=99,
  giveninits,
  uniquename=init]{biblatex} 
\usepackage{graphicx}
\usepackage{scrhack} 
\usepackage{listings}
\usepackage{lstautogobble}
\usepackage{tikz}
\usepackage{pgfplots}
\usepackage{pgfplotstable}
\usepackage{booktabs} 
\usepackage[final]{microtype}
\usepackage{caption}
\usepackage[hidelinks]{hyperref} 
\usepackage{ifthen} 
\usepackage{pdftexcmds} 
\usepackage{paralist} 
\usepackage{subfig} 
\usepackage{siunitx} 
\usepackage{multirow} 
\usepackage{array} 
\usepackage{makecell} 
\usepackage{pdfpages} 
\usepackage{adjustbox} 
\usepackage{tablefootnote} 
\usepackage{threeparttable} 

\usepackage{amssymb}
\usepackage{pifont}

\usepackage[acronym,xindy,toc]{glossaries} 
\makeglossaries

\newacronym[shortplural={D$_{\text{dye}}$}, longplural={donor dye, ex. Alexa 488}]{ddye}{D$_{\text{dye}}$}{donor dye, ex. Alexa 488}

\newacronym[description={\glslink{r0}{F\"{o}rster distance}}]{R0}{$R_{0}$}{F\"{o}rster distance}

\newglossaryentry{r0}
{
  name=\glslink{R0}{\ensuremath{R_{0}}},
  text=F\"{o}rster distance,
  description={F\"{o}rster distance, where 50\% ...}, 
  sort=R
}

\newglossaryentry{kdeac}
{
  name=\glslink{R0}{\ensuremath{k_{DEAC}}},
  text=$k_{DEAC}$, 
  description={is the rate of deactivation from ... and emission)}, 
  sort=k
}

\newacronym[shortplural={TUM}, longplural={Technical University of Munich}]{tuma}{TUM}
{Technical University of Munich}

\newglossaryentry{tum}
{
  name = TUM,
  description = {A university in Germany, Bavaria, in the city of Munich},
  plural = TUM
}

\newglossaryentry{computer}
{
  name=computer,
  description={is a programmable machine that receives input,
               stores and manipulates data, and provides
               output in a useful format}
} 

\bibliography{bibliography}

\setkomafont{disposition}{\normalfont\bfseries} 
\BeforeTOCHead[toc]{{\cleardoublepage\pdfbookmark[0]{\contentsname}{toc}}}

\definecolor{TUMBlue}{HTML}{0065BD}
\definecolor{TUMSecondaryBlue}{HTML}{005293}
\definecolor{TUMSecondaryBlue2}{HTML}{003359}
\definecolor{TUMBlack}{HTML}{000000}
\definecolor{TUMWhite}{HTML}{FFFFFF}
\definecolor{TUMDarkGray}{HTML}{333333}
\definecolor{TUMGray}{HTML}{808080}
\definecolor{TUMLightGray}{HTML}{CCCCC6}
\definecolor{TUMAccentGray}{HTML}{DAD7CB}
\definecolor{TUMAccentOrange}{HTML}{E37222}
\definecolor{TUMAccentGreen}{HTML}{A2AD00}
\definecolor{TUMAccentLightBlue}{HTML}{98C6EA}
\definecolor{TUMAccentBlue}{HTML}{64A0C8}

\pgfplotsset{compat=newest}
\pgfplotsset{
  cycle list={TUMBlue\\TUMAccentOrange\\TUMAccentGreen\\TUMSecondaryBlue2\\TUMDarkGray\\},
}

\graphicspath{
    {logos/}
    {figures/}
}

\newcolumntype{P}[1]{>{\centering\arraybackslash}p{#1}} 
\newcolumntype{M}[1]{>{\centering\arraybackslash}m{#1}} 
\newcolumntype{L}[1]{>{\raggedright\arraybackslash}m{#1}} 
\newcolumntype{R}[1]{>{\raggedleft\arraybackslash}m{#1}} 

\usepackage[ruled,vlined]{algorithm2e}

\usepackage{algorithm2e}
\SetKwFor{ForP}{Proof}{}{fine per}%
\SetKwFor{ForT}{Theorem}{}{fine per}%
\SetKwFor{ForS}{Statement}{}{fine per}%
\SetKwFor{ForL}{Lemma}{}{fine per}%

\newcommand*{\getUniversity}{Technische Universität München}
\newcommand*{\getFaculty}{Department of Informatics}
\newcommand*{\getTitle}{Probabilistic Population Protocol Models}
\newcommand*{\getTitleGer}{Probabilistische Modelle von Population Protocols}
\newcommand*{\getAuthor}{Vladyslav Melnychuk}
\newcommand*{\getDoctype}{Bachelor's Thesis in Informatics}
\newcommand*{\getSupervisor}{Prof. Dr. Javier Esparza}
\newcommand*{\getAdvisor}{Prof. Dr. Mikhail Raskin}
\newcommand*{\getSubmissionDate}{15.07.2022}

\newtheorem{theorem}{Theorem}
\newtheorem{definition}[theorem]{Definition}
\newtheorem{lemma}[theorem]{Lemma}
\newtheorem{statement}[theorem]{Statement}
\newtheorem{corollary}[theorem]{Corollary}

\begin{document}


\selectlanguage{english}

\pagenumbering{alph}

\frontmatter{}

\begin{titlepage}

\begin{center}

\includegraphics[width=40mm]{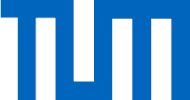}

  \vspace{5mm}
  {\huge\MakeUppercase{\getFaculty{}}}\\

  \vspace{5mm}
  {\large\MakeUppercase{\getUniversity{}}}\\

  \vspace{20mm}
  {\Large \getDoctype{}}

  \makeatletter
  \vspace{15mm}
  \ifthenelse{\pdf@strcmp{\languagename}{english}=0}
  {
  {\huge\bfseries \getTitle{}}

  \vspace{10mm}
  {\huge\bfseries \foreignlanguage{ngerman}{\getTitleGer{}}}
  }
  {
  {\huge\bfseries \getTitleGer{}}

  \vspace{10mm}
  {\huge\bfseries \foreignlanguage{english}{\getTitle{}}}
  }
  \makeatother

  \vspace{15mm}
  \begin{tabular}{l l}
    Author:          & \getAuthor{} \\
    Supervisor:      & \getSupervisor{} \\
    Advisor:         & \getAdvisor{} \\
    Submission Date: & \getSubmissionDate{} \\
  \end{tabular}

  \vfill{}
  
\end{center}
\end{titlepage}

\chapter{\abstractname}


Population protocols are a relatively novel computational model in which very resource-limited anonymous agents interact in pairs with the goal of computing predicates. We consider the probabilistic version of this model, which naturally allows to consider the setup in which a small probability of an incorrect output is tolerated. The main focus of this thesis is the question of confident leader election, which is an extension of the regular leader election problem with an extra requirement for the eventual leader to detect its uniqueness. Having a confident leader allows the population protocols to determine the convergence of its computations. This behaviour of the model is highly beneficial, and was shown to be feasible when the original model is extended in various ways \cite{michail2015terminating, jaax2020population, aspnes2017clocked, blondin2019expressive}. 

We show that it takes a linear in terms of the population size number of interactions for a probabilistic population protocol to have a non-zero fraction of agents in all reachable states, starting from a configuration with all agents in the same state. This leads us to a conclusion that confident leader election is out of reach even with the probabilistic version of the model.

\makeatletter
\ifthenelse{\pdf@strcmp{\languagename}{english}=0}
{\renewcommand{\abstractname}{Kurzfassung}}
{\renewcommand{\abstractname}{Abstract}}
\makeatother

\chapter{\abstractname}

\begin{otherlanguage}{german} 

Population protocols sind ein relativ neues Rechenmodell, bei dem sehr ressourcenbegrenzte anonyme Agenten in Paaren mit dem Ziel interagieren, Zahlenprädikate zu berechnen. Wir betrachten die probabilistische Version dieses Modells, die auch die Möglichkeit bietet die Situationen zu betrachten, in denen eine kleine Wahrscheinlichkeit einer falschen Ausgabe toleriert wird. Der Hauptschwerpunkt dieser Arbeit liegt auf der Frage der zuversichtlichen "leader election", einer Erweiterung des herkömmlichen "leader election"-Problems mit der zusätzlichen Anforderung, dass der endgültige "leader" seine Einzigkeit erkennen kann. Ein zuversichtlicher "leader" ermöglicht es population protocols, die Konvergenz ihrer Berechnungen zu bestimmen. Dieses Verhalten des Modells ist sehr vorteilhaft und es wurde gezeigt, dass es erreichbar ist wenn das ursprüngliche Modell auf verschiedene Weisen erweitert wird \cite{michail2015terminating, jaax2020population, aspnes2017clocked, blondin2019expressive}.

Wir zeigen, dass ein probabilistisches population protocol eine lineare Anzahl von Interaktionen in Bezug auf die Populationsgröße benötigt, um einen von Null verschiedenen Anteil von Agenten in allen erreichbaren Zuständen zu haben, ausgehend von einer Konfiguration, in der sich alle Agenten im gleichen Zustand befinden. Dies führt uns zu der Schlussfolgerung, dass eine zuversichtliche "leader election" selbst mit der probabilistischen Version des Modells unerreichbar ist.

\end{otherlanguage}

\makeatletter
\ifthenelse{\pdf@strcmp{\languagename}{english}=0}
{\renewcommand{\abstractname}{Abstract}}
{\renewcommand{\abstractname}{Kurzfassung}}
\makeatother
\microtypesetup{protrusion=false}
\tableofcontents{}
\microtypesetup{protrusion=true}

\mainmatter{}


\chapter{Introduction}\label{chapter:introduction}

Imagine a fictional village, which residents have just found out that they are about to be attacked by another, much stronger settlement. While each of the villagers has its own opinion on whether to surrender or heroically fight, they are all scared. So, they are running around in panic, while briefly exchanging their standpoint with the people they bump into, in order to find out what does the majority stand for. This is one of the scenarios that can be simulated with population protocols - a model of distributed computation introduced by Anguin et al. in 2004 \cite{ang2004}. Since then, it has gained a lot of interest in the research community, in fact, the updated version of the original paper \cite{ang2006} has been cited 694 times at the moment of writing this thesis. In this model, a collection of identical finite-state agents with very limited computational resources interact in pairs. In the standard setup, scheduling of such interactions is adversarial, but still subject to some fairness constraints. Namely, the global fairness condition guarantees that if a certain population configuration is reachable infinitely often, it is also reached infinitely often.  

The population protocols model and its variations found their application in several areas including engineering of sensor networks, chemical kinetics \cite{shah2019implementing} and swarm robotics \cite{shlyakhov2017survey, dudek2000computational}. For example, in order to describe various chemical computations, a closely related model of chemical reaction networks (CRNs) is used \cite{cummings2016probability}. In 2018, M. Vasic et al. have even introduced a new programming language CRN++ which allows to engineer artificial chemical systems and perform computations on them \cite{crn2018}.

Throughout the run of a standard population protocol, network agents interact with the end goal of stably computing a boolean function (i. e. predicate) on the initial states. This means that a some point of time, the computation converges with all the agents having a common correct value of the predicate. Even though individual agents can only store a constant number of bits in their states, and their computational ability is restricted to updating the state upon interacting with each other, the population as a whole can solve a lot of non-trivial problems. 

In this work we will consider probabilistic population protocols that allow a small probability of error in their computations. Turns out, that in the probability $<1$ setting, population protocols, as well as CRNs, can with high probability simulate a LOGSPACE Turing machine (and a register machine respectively), and hence decide Turing-computable predicates \cite{ang2006, cummings2016probability} under one additional assumption. The last one is quite simple, and, in fact, used in the majority of prior studies: it is assumed that the scheduler selects the ordered pair of agents to interact independently and uniformly at random among all possible ordered pairs corresponding to edges in the complete interaction graph. 

In this thesis, we investigate the question of existence of a probabilistic protocol that could solve one of the most common computational tasks for this model - the leader election problem, in a way that the elected leader can eventually detect its uniqueness.


\section{Other Related Work}

Since the appearance of population protocols, this model has been studied from many different perspectives, including its expressive power, verification complexity, impact of adding randomness to the model, optimal stabilization time for particular computational tasks, etc.  

\vspace{1mm}
\noindent
\textbf{Expressiveness and verification complexity.} In has been shown that the standard model is capable of computing exactly the semilinear predicates, which correspond to precisely the predicates definable in first-order Presburger arithmetic \cite{presburger, ang2007}. Although the expressive power of the model had been exhaustively studied in the first years after its introduction \cite{ang2006, ang2007}, the correctness problem for population protocols (i. e. whether the protocol reaches the right consensus for all inputs) has only been shown to be decidable in 2015 by Esparza et al. \cite{esparza2015verification}.

\vspace{1mm}
\noindent
\textbf{Model variations.} A lot of prior studies introduced extensions and restrictions to the original model. For example, Anguin et al. studied the expressive power of the \textit{one-way communication} model, in which the communication between agents is asynchronous \cite{ang2007}. This restriction of the communication structure led to further variations of the population protocol model, such as \textit{transmission}, \textit{observation}, \textit{immediate delivery}, \textit{delayed delivery}, and \textit{queued delivery} models. None of these models appeared to be more expressive than the standard model with two-way communication.

In this thesis we in particular look at the problem of detecting the convergence of the computation by the population protocol. To our knowledge, none of the prior studies have extensively approached this problem in terms of a standard population protocol model. On the other hand, several works have introduced extensions to the model in order to make this useful behaviour accessible. All of the following population protocol model extensions allow to detect the termination of a computation:

\begin{itemize}
    \item Michail and Spirakis considered the population protocols equipped with a \textit{cover-time service} which is given by a special state that can store the event of his interaction with every agent in the whole network \cite{michail2015terminating}. This model appears to have the computational power between SPACE(log $n$) and NSPACE(log $n$). They also show that this model is reducable to another model variation capable of performing \textit{halting computations} - \textit{Population Protocols with Absence Detector}.
    \item Jaax looked at the assumption of the global knowledge of the population size in a population protocol \cite{jaax2020population}.
    \item Aspnes introduced \textit{clocked protocols}, that argument the standard model with a clock oracle that signals to an agent when it has waited long enough for the protocol to have converged \cite{aspnes2017clocked}. These protocols have a power equivalent to NL.
     \item Blondin et al. studied \textit{broadcast protocols} which extend the population protocols with reliable signal broadcasts and showed that this variation is capable of computing precisely the predicates in NL. \cite{blondin2019expressive}. 
\end{itemize}

As we see, being able to detect the convergence of a computation, allows the protocols to compute more difficult tasks than those described by just semilinear predicates. 

\vspace{1mm}
\noindent
\textbf{Leader election.} Leader election is one of the most common computational tasks in population protocols field, that we also focus on in this thesis. In essence, in the beginning of the computation each agent is equipped with a leader bit, which can be either $0$ or $1$. The goal is to reach a configuration, in which only one agent has a leader bit equal to $1$, i.e. is a single "leader". Additionally, for the agent to be a confident leader, means knowing (with high probability) that he is the only remaining agent with leader bit equal to $1$. 

No wonder that a lot of works study this question, especially in terms of the optimal expected time for the protocol to elect a leader. The optimal stabilization time appears to strongly vary with the number of protocol states allowed \cite{doty2018stable, alistarh2017time, sudo2020leader}. In this thesis we stick to the constant amount of states, and in this case a linear parallel time, in other words, $O(n^2)$ transitions are required for a protocol solving the leader election problem to stabilize \cite{ang2006}.

\vspace{1mm}
\noindent
\textbf{Probabilistic version of the model.} The original paper by Angluin et al. also introduced a natural probabilistic variation of the population protocol model in form of a model of \textit{conjugating automata} \cite{ang2004}. This ultimately allowed to study problems where the correct output is computed with probability smaller than $1$. They have also shown, by using an epidemic-based \textit{phase clock mechanism} (that can be implemented in the probabilistic protocols model) that a population with an initial leader can in fact with high probability simulate a register machine, and as result detect that a certain computation is likely to have converged \cite{angluin2008fast}. As we will see a bit later, the similar techniques of high probability time-passage detection fail to solve the Confident leader election problem. The existence of a population protocol that is able to solve this problem was stated as an open question in \cite{doty2018exact}, which leads us to the contributions of our work.

\section{Contributions of this Work}

In this thesis we consider the low-error-probability population protocols and formalize the Confident leader election problem, which to our knowledge, was not explicitly done in prior studies.

We show that for any such population protocol with a constant number of states there exists a point of time, such that each reachable state is occupied by $\theta(n)$ agents asymptotically almost surely for $n\rightarrow \infty$. As a corollary, we show the impossibility of confident leader election in the low-error-probability population protocols, which is our main contribution. We also present the results of simulations that led us to and support our results.

Apart from that, we formally describe the leader election construction by Angluin et al. \cite{ang2006} that was used in their proof of the statement that probabilistic population protocols can efficiently simulate LOGSPACE Turing machines. We argue with the help of simulations that their method, as well as other timer- and clock-based techniques are not applicable for solving the Confident leader election problem, which led us to our main result stated above.

\section{Importance of Results}

One of the co-authors of the original paper on population protocols, J. Aspnes, argues in one of his recent papers \cite{aspnes2017clocked}, that the inability to detect convergence of the computations is the fundamental limitation of the standard population protocols model, which is the reason its power is limited to just computing the predicates definable in Presburger arithmetic. While allowing for a small error may give the model additional computational power, to our knowledge, none of the previous research has considered the question of confident leader election in this setup. 

Having a confident leader gives the model ability to notify the outside observer about the end of the computation, which is an expected part of the modern programming languages. In case of a multi-stage computation, it would be possible to keep track of its advances, making the computation more predictable and easier to analyze. Of course, all this would come at a cost of a time overhead, but in many cases that would be negligible in comparison to the main computation.

Motivated by the above advantages, we were initially convinced that it is possible to find a procedure that would solve the confident leader election. In essence, one would just have to implement an instrument for tracking time, so that after waiting long enough one would conclude with high probability that a single confident leader is left. As we noted earlier, this idea is not new and has been previously realized for similar purposes by using a phase clock mechanism or just by marking the agents as "timers" \cite{ang2006, aspnes2017clocked, angluin2008fast}, and hence, seemed very promising. Only after running a range of simulations it became clear, that this tempting result is most likely out of reach.

\section{Outline}

This thesis is structured as follows. In Chapter 2, we define the formal model of population protocols and formally describe how its performs computations. We also introduce probabilistic population protocols and their essential component - probabilistic model of agent selection. Chapter 3 presents the definitions of correctness and correctness probability in the population protocols, which leads us to formalizing the notion of asymptotically almost surely correct population protocols. In this chapter we also introduce the main subject of interest of this thesis - the Confident leader election problem. We then discuss in detail the leader election protocol used by Angluin et al. in \cite{ang2006} and present its slightly improved version. In the end of Chapter 3, we show the simulation results that give insights in how these protocols operate in practise. Chapter 4 is the central part of this work, as it states our main contribution and provides its proof. In Chapter 5 we present one more interesting simulation experiment which leads to the main open questions. Finally, Chapter 6 gives a summary of achieved results and gives an outlook on future work.

\chapter{Preliminaries}\label{chapter:preliminaries}

Before getting into the formal definitions, let us first explain what the essential parts of the population protocol are, and how it performs computations. A population protocol runs on a population, consisting of anonymous agents. Each agent starts in an initial state, which is represented as bits of information that the agent stores in its limited memory. During the run of the protocol, at each time step a scheduler picks the ordered pair of agents to interact. Upon interacting, the agents change their states (possibly keeping the same state). This changes are determined by the states of both agents before the interaction and the transition function. The mapping of agents to their states forms the population configuration. Note, that since the agents in the same state are indistinguishable, the number of agents in each state exhaustively describes the state of the population as a whole too. The canonical property of the scheduler is to satisfy the fairness condition. Informally, it states that each reachable configuration is reached infinitely many times. 

The output of the computation is determined with a help of a predefined function, that assigns each state a value in $\{0,1\}$. This way each configuration can be mapped to an output assignment. We say that the computation stabilizes, if the output assignment of the current configuration cannot be changed with future interactions. This builds the foundation for the stable computation of predicates, which we will also discuss in this chapter.

\section{Formal Model}

We introduce population protocol model, closely following the initial definitions of Anguin et al. \cite{ang2006}.

\begin{definition}
    A population protocol \(\mathcal{A}\) is a tuple $(Q, \Sigma, I, O, \delta)$, consisting of the following components:
    \begin{itemize}
        \item $Q$, a finite nonempty set of possible agent states,
        \item $\Sigma$, a finite nonempty input alphabet,
        \item $I:\Sigma \rightarrow Q$, an input mapping function, with $I(\sigma)$ denoting the initial state of an agent with input $\sigma$,
        \item $O:Q\rightarrow \{0, 1\}$, an output mapping function, with $O(s)$ denoting the output value of an agent in state $s$, and
        \item $\delta: Q \times Q \rightarrow Q \times Q$, a transition function on pairs of states.
    \end{itemize}
\end{definition}

The output mapping function $O$ can generally have values other than $0$ and $1$. This would not make any difference for our main proofs, so we kept the definition simple.

In order to show that the interaction between agents in states $s_1$ and $s_2$ changes their states to $s_1'$ and $s_2'$ respectively, we will write $\delta(s_1,s_2)=(s_1',s_2')$. 

\begin{definition}
    A population \(\mathcal{P}\) is a pair $(A, E)$, where
    \begin{itemize}
        \item $A$ is a finite set of $n$ agents forming the population, and
        \item $E \subseteq A \times A$ is an irreflexive relation representing the pairs of agents that can interact with each other, i.e. the edges of the interaction graph.
    \end{itemize}
\end{definition}

In this thesis, as in the majority of prior studies, we will only consider the situation, when the interaction graph is complete, meaning that all possible pairs of agents can interact. Note, that in the population protocol model the number of agents is fixed (unlike, for example, in CRNs \cite{cummings2016probability}). 

\begin{definition}
    A population configuration is a mapping $C : A \rightarrow Q$, with $C(a)$ denoting the state of an agent $a$.
\end{definition}

We call the state $s$ $empty$ if $\forall a \in A:C(a)\neq s$, where $C$ is the current configuration in the run of the population protocol. If configuration $C'$ can be reached from configuration $C$ in one transition, we write $C\rightarrow C'$.

\begin{definition}
    An execution sequence is a (possibly infinite) sequence of configurations $(C_1,C_1,...)$, such that $\forall i:C_i \rightarrow C_{i+1}$. For each execution sequence there exists a corresponding transition sequence, from which the execution sequence was induced. 
\end{definition}

The lengths of execution and transition sequences are naturally the number of configurations and transitions in them respectively. 

\begin{definition}
    \label{reachable}
    A configuration $C$ is called reachable if there exists a finite execution sequence containing $C$ and starting with the initial configuration of the protocol $C_0$. A state $s$ is called reachable if there exists a reachable configuration $C$, for which $s$ is not empty.
\end{definition}

The main purpose of the population protocols is to carry out computations, to be more precise - compute predicates. At each point of time, the output mapping function $O$ assigns an output value to each state in a configuration, and therefore - to each agent. In case if all the non-empty states have the same output, we say that the agents agree on the output, or the configuration is a $consensus$.

\begin{definition} (Adapted from \cite{raskin2021population}).
    A configuration $C$ is called a stable consensus if it is a consensus, and for each agent a the output of its state does not change in any configuration $C'$ that is reachable from $C$.
\end{definition}

Those population protocols that reach a stable consensus on every input $I$, and the output value of this consensus for each input $I$ is always the same, $stably$ $compute$ a predicate $F:X\rightarrow \{0,1\}$ in a natural way, where $X$ is the set of all possible initial assignments of agents to their input. Most of the studies mainly focus on this type of protocols. Going forward, in this paper we will also consider protocols that do not necessarily fulfill this property. 

\section{Probabilistic Model of Agent Selection}

The global fairness condition on the scheduler is the key to analyzing the complexity of predicates computable in the population protocol model. This condition is met, if each infinite execution sequence chosen by the scheduler is $fair$.

\begin{definition}
    An infinite execution sequence $(C)$ is fair if for each configuration $C'$, such that $C\rightarrow C'$ and $C$ occurs infinitely in $(C)$, $C'$ also occurs infinitely often in $(C)$. 
\end{definition}

We call an infinite fair execution a $run$ of the population protocol. However, sticking solely to the fairness conditions does not allow to make any strong guarantees on the expected running time of the protocol before reaching a stable consensus. Hence, we follow the majority of previous research on population protocols and consider an additional natural probabilistic assumption of working with a \textit{uniformly random scheduler}.

\begin{definition} (Adapted from \cite{sudo2019logarithmic}).
    \label{def_uniform_scheduler}
    A uniformly random scheduler $\Gamma$ corresponds to an infinite sequence of interactions $\Gamma_0,\Gamma_1,...$, where each $\Gamma_t$ is a random variable with $Pr[\Gamma_t=(a,b)]=\frac{1}{n(n-1)}$ for any $t\geq0$, where $(a,b)$ is an ordered pair to interact of any distinct $a,b \in A$. 
\end{definition}

The key to show that the sequences of interactions from Definition \ref{def_uniform_scheduler}, are fair with probability $1$, is that every configuration in a population protocol model has a finite amount of successors alongside with the fact that each possible pair of agents is scheduled as the next pair to interact with a positive probability. A formal proof of why probabilistic protocol model (which we will introduce in the next section) meets the fairness condition when using a uniformly random scheduler can be found in Section $7.3$ of the paper by Esparza et al. \cite{esparza2021complexity}.

Apart from being able to study the expected time until the computation stabilizes, the major advantage of this probabilistic model of agent selection is that we can now consider situations, in which the protocol does not necessarily always have the same output on a certain input upon stabilizing. This means that we can study the protocols, which correctly compute predicates with probability less than 1.

\section{Probabilistic Population Protocols}

\begin{definition}
 A population protocol is probabilistic, if each configuration $C$ creates a probability distribution over the set of directly reachable configurations $\{C'|C \rightarrow C'\}$.
\end{definition}

From now and on, we will work with probabilistic population protocols that use a uniformly random scheduler, unless explicitly stated otherwise. Now we have the necessary preliminaries to consider an error-prone computation with population protocols, and calculate the expected convergence time for executions.




\chapter{Low-error-probability Population Protocols}\label{chapter:low_error_ppp}

For many computational models in makes sense to consider the probabilistic version of the model that does not always yield the correct output. In this chapter, we formalize the behavior of low-error-probability population protocols and present the leader election problem. Leader election is a crucial procedure that is a part of computing many basic, but also non-trivial predicates. The last ones can be often viewed and computed as sequences of simple operations. For this reason, it is highly beneficial for the protocol to know when to move one from one operation to another. The most intuitive solution here would be to implement a procedure that will notify all agents about the fact that the computation of the current operation has finished with high probability. This, in its core, comes to a confident leader election, which is the main focus of this thesis. 

Anguin et al. followed a similar approach of breaking a complex function into simple operations, when showing that probabilistic population protocols lie in randomized LOGSPACE \cite{ang2006}. Yet they compute it in an alternative way, omitting the confident leader election. We will explain their method of leader election and discuss it.  


\section{Correctness Probability and Formal Model}

Before we start working with error-prone protocols, it is important to give a proper definition of correctness and correctness probability. Note, that intuitively these definitions only make sense subject to some particular computational task. In the following, we assume that we can determine whether a certain output is correct for a given input just by using the definition of such a task. For example, in the case of leader election (which we introduce shortly) the "correctness measure" of a configuration is to have exactly one agent is the "leader" state.

\begin{definition}
    \label{def_corr}
    For a population protocol \(\mathcal{A}\) and a computational task $F$, we say that a configuration $C$ of \(\mathcal{A}\) is correct if the output of each agent in $C$ is correct subject to $F$ and the given input $I$. $C$ is stably correct, if additionally every configuration reachable from $C$ is correct. 
\end{definition}

Notice that we wanted to make Definition \ref{def_corr} not specific to just predicates, therefore we formulated it in a more general form than just by using the previously introduced notions of consensus and stable consensus.

\begin{definition}
    A population protocol \(\mathcal{A}\) stably computes a computational task $F$ with probability p, if, with probability p, it reaches a stably correct configuration. If p=1, we just say that a \(\mathcal{A}\) stably computes $F$.
\end{definition}

Note, that we do not specify what happens if a stably correct computation is never reached. This is because we do not require the protocol to stabilize in this case.

The last definition leads us to the notion of low-error-probability population protocols, which fulfil $p>0.5$. We also find it useful to define the following special sub-category of low-error-probability protocols.

\begin{definition}
    For a given computational task $F$, an asymptotically almost surely (a.a.s) correct population protocol, is a population protocol that stably computes $F$ with probability $p=1-o(1)$ w.r.t. n, where n is the population size.
\end{definition}

A.a.s. correct population protocols are of a great practical interest, since as the problem instances get larger, the probability of correct computation approaches $1$, even without the overhead of success amplification methods.

\section{Leader Election}

Leader election is one of the most common problems in the population protocols, which has been exhaustively studied with different additional constraints \cite{doty2018stable, alistarh2017time, sudo2020leader}. The \textit{Leader election problem} can be formulated as follows:

\vspace{5mm}
\textit{Construct a population protocol \(\mathcal{A}\), that has two \textbf{types} of states: L and F (standing for leader and follower) - meaning that each state includes a flag bit denoting whether the state is of the "leader" type or not, and:
\begin{enumerate}
    \item Each possible initial configuration reaches a configuration $C$, in which only one agent is in the state of type L and all the others - in states of type F;
    \item For every configuration $C'$ reachable from $C$ it holds: $\forall a \in A: stateType_C(a)=stateType_{C'}(a)$, where $stateType_C(a)$ denotes the state type of the agent $a$ in configuration $C$. 
\end{enumerate}}
\vspace{5mm}

In this work we only consider protocols with a constant amount of states. It was shown, that in this case the best expected parallel time to stabilize for a population protocol solving the Leader election problem is $O(n)$ \cite{doty2018stable}.  


As we noted earlier, for the reasons of enabling the composition of complex predicates out of simple operations and being notified when the computation stabilizes with high probability, it is useful for the eventual leader to detect its leadership. A few studies have considered the confident leader election before \cite{aspnes2017clocked, doty2018exact}, but to our knowledge, none of them gives a formal definition of what being a confident leader actually means. Before we give a natural definition for it, let us make a following observation. The word "confident" here essentially means that an agent believes that it is the \textbf{only} leader remaining. It means that if another agent, possibly after some period of time also becomes a confident leader, both of them have failed to detect each other. For this reason, there must exist an exclusive state which will be only visited by one agent, namely a confident leader. Now, let us put it in a more formal way.

\begin{definition}
    \label{conf_leader}
    The Confident leader election problem is a problem of designing a population protocol \(\mathcal{A}\) that firstly solves the Leader election problem and additionally after that reaches a configuration $C''$ in which the elected leader agent is in a state of a special \textbf{sub-type} CL of type L (which stands for "confident leader"), all the other agents are in states of type F, and: 
    \begin{itemize}
        \item for every configuration $C'''$ reachable from $C$ it holds: $\forall a \in A: stateType_{C''}(a)=stateType_{C'''}(a)$, where $stateType_C(a)$ denotes the state type of the agent $a$ in configuration $C$;
        \item only one agent, namely the elected leader after solving the Leader election problem, reaches the state of a sub-type $CL$ throughout the whole protocol run.
    \end{itemize}
     An agent can only be in the sub-type K of type J, if this agent is in type J. This means that in our case an agent cannot be in a state of sub-type CL while also being in a state of type F.
    
\end{definition}

It is easy to show that it is impossible to construct a population protocol that solves the Confident leader election problem with probability $1$. Note, that the probability of two agents only having rendez-vous with each other agent during the upcoming $k$ interactions (in which at least one of them is present) is positive, and is equal to $(n-1)^{-k}$. This means, that for any fixed number of interactions there is a positive probability that these two agents are "isolated" from the other agents during these $k$ interactions. Since the two groups of agents have a chance of not exchanging any information, none of the agents can be certain that no other leader exists.

However, it is not clear what happens if we switch to the error-prone model.

\section{Using Counters to Detect the Termination of Leader Election}

In 2006, the authors of the population protocols model, Angluin et al., have shown that population protocols can simulate a logspace Turing machine with high probability \cite{ang2006}. We will give a brief overview of their method, and then focus on how their leader election procedure works. The central idea of the proof was to firstly simulate a counter machine with high probability, and then use the standard reduction due to Minsky \cite{minsky1967computation} from Turing machines to counter machines. The main functionality of the counter machine is to \textit{simulate counters} and perform \textit{zero tests}. Both rely on having a leader agent which could organize the rest of the population to behave in a certain way. 

The reason why we are interested in the leader election approach used in this proof, is that it is designed in a way, such that the leader knows when to move on to the next computational step, after being elected. This is very similar to our end goal of having a confident leader! The big difference, however, lies in the fact, that this leader election procedure allows error-correction, meaning that if two leaders meet after the next computational step has began, one of them (the winner) keeps the leader bit, and restarts the \textit{initialization phase}. Let us describe this protocol in more detail.

\vspace{1mm}
\noindent
\textbf{Protocol 1 } (Adapted from \cite{ang2006}). 
In addition to the actual received input (with is remembered throughout the whole computation), each agent stores the values for the following \textit{additional variables}: \textit{leader bit }$\in\{0,1\}$, \textit{timer bit }$\in\{0,1\}$, \textit{timer\_set bit }$\in\{0,1\}$, \textit{timer\_reset bit }$\in\{0,1\}$, \textit{timer\_count }$\in\{0,k\}$, where $k$ is a predefined value which we will describe a bit later. We call an agent with leader bit equal to $1$ a $leader$, and an agent with timer bit equal to $1$ a $timer$. In the beginning of protocol's run, every agent sets its leader bit to $1$, all the other additional variables to $0$, and proceeds to the initialization phase.

\vspace{1mm}
\noindent
\textbf{Initialization phase.}
Let $\delta$ be the transition function of the protocol. To show how the values of the additional variables change upon interaction, we will write transitions in a simplified form, assuming that nothing else changes, unless stated otherwise. So, we let the state to be described by the tuple $(\textit{leader bit }, \textit{timer bit }, \textit{timer\_set bit }, \textit{timer\_reset bit }, \textit{timer\_count})$. We will give a short explanation below most of the transitions.

\begin{itemize}
    \item[1.] $\delta((1,0,0,0,0) ,(0,0,0,0,0))=((1,0,1,0,0) ,(0,1,0,0,0))$
    \vspace{1mm}
    
    A leader initially attempts to mark the first non-timer agent it encounters as a timer. If this happens, the leader sets its timer\_set bit to 1, so that no more than one timer is marked by the same agent at a time. 
    \item[2.] $\delta((1,0,1,0,tc_1) ,(0,1,0,0,0))=((1,0,1,0,tc_1+1) ,(0,1,0,0,0))$
    \vspace{1mm}
    
    If a leader who has already marked a timer meets a timer, it increases its timer\_count, which stands for the number of consecutive interactions with timer agents. If the timer\_count reaches a threshold value $k$, the initialization phase for this agent is finished, and it moves on to the actual computation, which is the central idea of this protocol. 
    \item[3.] $\delta((1,0,ts_1,0,tc_1) ,(1,0,0,0,0))=((1,0,1,0,0) ,(0,1-ts_1,0,0,0))$
    \vspace{1mm}
    
    If two leaders meet, one of them becomes a "loser" and deletes its leader bit, while the other one retains its leader status, but sets its timer\_count to $0$, thereby restarting the initialization phase. In case if "winner" has not marked a timer yet, it marks loser as a timer and sets its timer\_set bit to $1$. 
    
    \item[4.] $\delta((1,0,ts_1,0,tc_1) ,(1,0,1,0,tc_2))=((1,0,ts_1,1,0) ,(0,0,0,0,0))$
    \vspace{1mm}
    
    In case if loser has already marked a timer, the winner sets its timer\_reset bit to $1$, and waits until it encounters a timer to turn it back to a non-timer, before proceeding (see next transition).
    \item[5.] $\delta((1,0,ts_1,1,0) ,(0,1,0,0,0))=((1,0,ts_1,0,0) ,(0,0,0,0,0))$
    \item[6.] $\delta((1,0,1,0,tc_1) ,(1,0,ts_2,1,0))=((1,0,1,0,0) ,(1,0,ts_2,1,0))$
    \vspace{1mm}
    
    Although an agent with timer\_reset equal to $1$ only changes its state when interacting with a timer, it is still able to reset the timer\_count of another agent if it is not timer itself.
    \item[7.] $\delta((1,0,1,0,tc_1) ,(0,0,0,0,0))=((1,0,1,0,0) ,(0,0,0,0,0))$
    \vspace{1mm}
    
    The timer\_count of an a leader is also reset to $0$ if it encounters an agent that is neither a leader or a timer. 
\end{itemize}

Otherwise, nothing happens protocol-wise. Also, we set all of the above transitions, except for $3.$ and $4.$, to be symmetrical, in the sense that they are also applicable, when we exchange the order of agents in them: $\delta(s_1 ,s_2)=(s_1' ,s_2') \Rightarrow \delta(s_2 ,s_1)=(s_2' ,s_1')$. As always, only one (possibly dummy) transition applies for a certain pair of agents. 

Note, that the original protocol theoretically allows the agent to be a leader and a timer as the same time. But because of the facts that only a leader can mark a timer; a non-leader never turns into a leader, and one of the leaders becomes a non-leader when two leaders interact, we assume that this is not the case.

\vspace{1mm}
\noindent
\textbf{Computation phase.}
Once a leader has encountered a timer $k$ times in a row, it begins the actual computation (e. g. simulating counters and performing zero checks). If this agent meets another leader during the computation phase, they both proceed with the transitions $3.$ and $4.$, depending on which one is applicable. The winner then restarts the initialization phase. Here it is important to note that the goal of the protocol is to perform a computation with a single leader. Multiple leaders create a mess, and this is why it is important to ensure the appropriate retrieval of all the agents back to their initial input (but not the additional variables) if two leaders meet while at least one of them is in the computation phase. Angluin et al. do not explicitly mention how this is achieved in their paper, but this can be easily done by ensuring that the leader that has just launched the computation phase starts "working" with the initial input values of the agents that it meets.

After an expected time of $\theta(n^2)$ interactions, only one leader remains. In the following we will see, that, before this happens, it is very likely that some leaders went back to the start of their initialization phase, after having finished it at least once.

The idea of using timer agents to detect the passage of time is quite common for the population protocols \cite{ang2006, aspnes2017clocked, angluin2008fast}. The question that one may ask at this point, is whether it is possible to use a similar technique to be fairy confident, that if the leader has proceeded to the computation phase, it will not meet another leader and restart the initialization, in other words, be a confident leader. But first, let us take a closer look at how Protocol $1$ performs in practice.

Apart from this leader election protocol, we implemented our own, improved version of it \cite{gitlab}. The motivation for this was the general chaos in the beginning of the initialization phase, which led to a large amount of timers, which, in its turn, increased the probability to meet one $k$ times in a row. Our goal was to make timers sparse, even if the leaders are abundant. This way the probability to be a single leader upon moving to the computation phase would intuitively decrease. For this, we equipped the agents with one more additional variable - a $has\_seen\_timer$ $bit$ which is set to $1$ if the agent has already interacted with a timer, regardless of its own state. The agent with a has\_seen\_timer bit equal to $1$ can also remove the leader bit from the leaders who have not seen a timer yet. 

\begin{figure}[htp]
    \centering
    \includegraphics[width=220pt]{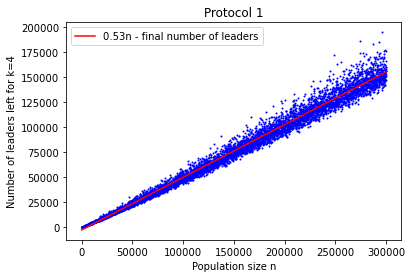}
    \includegraphics[width=220pt]{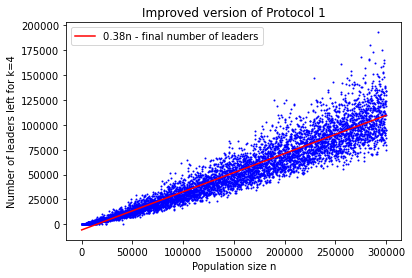}
    \caption{Simulation results: final number of leaders for k = 4}
    \label{fig:Sim_1}
\end{figure}

We ran the both protocols once for each value of $(100 + 50n)$ between $100$ and $300000$ for $n = 0,1,2,...$ up to the point when the first leader meets a timer $k$ times in a row, or there is only one leader left. We looked at the number of leaders still present in the population at this point of time - "final number of leaders". The results for $k=4$ are visualized in Figure \ref{fig:Sim_1}. 

\vspace{3mm}

We also ran the protocols for $k=3,5,6,7$ and observed a similar linear dependence between the final number of leaders and $n$. As we see, the modified version of the protocol shows a slight improvement in comparison to the original one. Nevertheless, it seems that no matter how high $k$ is chosen, the number of agents in the leader state at the point of time when the first leader proceeds to the computation phase grows linearly with $n$, and therefore, a confident leader election with no resets is likely to be impossible with these concrete protocols. 

Another example of a protocol that reduces the amount of leaders, starting in a configuration with every agent being a leader, is the Fast leader elimination protocol presented in 2008 \cite{angluin2008fast}. It is also based on the idea of tracking time, but using a \textit{phase clock} instead, which is a mechanism that in its core relies on the the number of consecutive coin tosses values equal to $1$ to signal the phase change. Although the correctness of this protocol is not explicitly proven, the authors provide a simulation that shows how the number of leaders is reduced by the end of the computation. It can be seen that this protocol generally leaves $\theta(n^{1-\epsilon})$ leaders, which, of course, is also not enough to produce a confident leader.

Multiple unsuccessful attempts of creating a leader election procedure that could determine when the convergence occurs, led us to the following conclusion. The reason why the clock- and timer-based protocols do not succeed in solving the Confident leader election problem is that such protocols rely on the fixed parameters, like $k$ in Protocol $1$, or the number of consecutive coin tosses or clock phases in the Fast leader elimination protocol. Hence, it is most likely impossible to track the passage of a non-constant amount of time, even with high probability. 

It turns out, that confident leader election is actually impossible with any probabilistic population protocol. In the next chapter, we present a formal proof of this statement.

\chapter{Main Impossibility Result}\label{chapter:main_result}

\section{Main Claim}

Let us first give the statement of our main theorem.

\begin{theorem}
    \label{main_theorem}
    Let \(\mathcal{A}\)$=(Q, \Sigma, I, O, \delta)$ be a population protocol with $O(1)$ states running on a population \(\mathcal{P}\)$=(A, E)$, where all agents start in the same state $s_0$. Then there exists a point of time $I$, such that at $I$ there are $\theta (n)$ agents at each reachable state of \(\mathcal{A}\) asymptotically almost surely w. r. t. $n\rightarrow \infty$, where $n$ is the population size.
\end{theorem}

From now and on, by $a.a.s.$ we will mean "asymptotically almost surely w. r. t. $n\rightarrow \infty$".

The statement of Theorem \ref{main_theorem} was inspired by the fact, that if the confident leader election was in fact impossible, a protocol could not contain a state that is only reached by a single agent throughout the protocol run with high probability. Since the transitions can be interpreted as "ratios" of the agents moving from one state to another, reaching a new state by one agent, will most likely mean that many other agents have a chance of reaching it by taking the same transitions. This led us to the assumption that eventually $\theta(n)$ agents will reach every reachable state.

\section{Proof Structure}

Since the structure of our proof of Theorem \ref{main_theorem} is not trivial, we aim to give the reader its overview in this section. It is visualized on Figure \ref{proof_structure}. The proof starts with a mathematical-induction-friendly reformulation of the main theorem - Statement \ref{main_ind}. Essentially, we view the history of interactions as an "interaction forest", and we start from the point where all agents are in the same state and add "layers" to it one by one. Each of these layers is a set of states that are reachable in $k$ steps, but not reachable in $k-1$ or less steps. This gives us the ability to prove the existence of a point of time with all the states of the current "$(k+1)$-layered" forest having $\theta(n)$ agents by assuming the correctness of this statement for the "$k$-layered" forest.   

It is a lot easier to show the inductive step of Statement \ref{main_ind} by looking at one state at a time, so we use another mathematical induction and Lemma \ref{second_ind} to make this happen. In the inductive step of Lemma \ref{second_ind}, in its turn, we need to prove that if at a certain point of time we have $\theta(n)$ agents in all of the states of some set $S$, and there exists a transition on two of these states which allows the agent to move to a state which is not in $S$, then after some time this new state, alongside with all the states from $S$, will be populated by $\theta(n)$ agents. We split this statement in two parts and show them with Lemma \ref{marking} and Lemma \ref{untouched} - they are the heart of our proof. Defining a variety of random variables and additional procedures, allowed us bound the probabilities of occurrence of some interesting events by using well-known distributions and inequalities. 

After showing the inner-most induction, it just remains to carefully finish the corresponding proofs and close the chain of implications that leads us to Theorem \ref{main_theorem}. 

\DontPrintSemicolon

\begin{figure}[h]

    \begin{algorithm}[H]
    \SetAlgorithmName{Proof Structure}{}
    \textbf{Theorem \ref{main_theorem}}\;
    \ForP{}{
        \textbf{Statement \ref{main_ind}}\;
        \ForP{}
        {
        Base step\;
        Induction hypothesis\;
        Induction step\;
          \ForP{}
            {
            \textbf{Lemma \ref{second_ind}}\;
              \ForP{}
                {
                    Base step\;
                    Induction hypothesis\;
                    Induction step\;
                    \ForP{}
                        {
                            \textbf{Lemma \ref{marking}}\;
                            \ForP{}
                                {
                                    
                                }
                            \textbf{Lemma \ref{untouched}}\;
                            \ForP{}
                                {
                                    \textbf{Statement \ref{statement_untouched}}\;
                                        \ForP{}
                                            {
                                                
                                            }
                                }
                        }
                        
                }
            }
        }
    }
    \caption*{}
        
\end{algorithm}
\caption{Proof Structure}
\label{proof_structure}
\end{figure}

\section{Proof of Main Result}

The central goal of our thesis to show that high probability confident leader election is not feasible with standard population protocols. Due to Definition \ref{conf_leader}, a confident leader must be the only agent reaching a special state of sub-type $CL$ throughout the whole protocol run. Our idea lies in showing that the existence of a state that is only visited by one agent is a.a.s. impossible in the first place. Before getting into the proof, we will formalize the exact procedure of how the next pair of agents to interact is selected.

\textbf{ChooseNextPair.} As we mentioned earlier, usually in population protocols the fairness of the model is reached through the assumption that the ordered pair to interact is chosen uniformly at random among all possible ordered agent pairs \cite{ang2006}. We will define the most intuitive procedure to get such a pair of agents and we will make use of it throughout the whole proof:

\vspace{5mm}
\begin{algorithm}[H]
    \caption{ChooseNextPair}
    \begin{itemize}
        \item[1.] Randomly choose agent $a$ from the whole agent set $A$.
        \item[2.] Randomly choose agent $b$ from the whole agent set $A$.
        \item[3.] If $a$ and $b$ are distinct, $(a,b)$ is the next pair of agents to interact. Otherwise, nothing happens protocol-wise. 
    \end{itemize}
\end{algorithm}
\vspace{5mm}

Due to the symmetry, this procedure indeed guarantees the uniform randomness of each chosen pair. In the following, we will assume that all the interactions are determined by it. This, essentially, leaves us with a uniformly random scheduler, that, in its turn, guarantees the global fairness.

\vspace{1mm}
\noindent
\textbf{Remark.} Throughout the proof, we will work with mathematical expressions denoting specific fractions of all agents that are in particular state. Naturally, these fractions will not correspond to natural numbers of agents for every $n$, but rather need to be rounded to become such. Our proof is robust to integer rounding in any direction (meaning that rounding retains the asymptotic behaviour of all the expressions), so, in order not to unnecessarily overcomplicate the computations, we will assume that these values are always rounded in one of the directions.

Now we have all the preliminaries needed for our proof.

\vspace{1mm}
\noindent
\textit{Proof of Theorem \ref{main_theorem}.} Let $s_0$ be the starting state, and let $F_0 = \{s_0\}$. Further, let $F_{i+1}=F_i \bigcup \{ s \in Q |\text{ }\delta(p,q)=(p', s) \vee \delta(p,q)=(s, q'), p,q \in F_i \}$. In other words, $F_i$ denotes the set of states reachable in less or equal to $i$ interactions from the starting configuration.

We will start with the following statement.

\begin{statement}
\label{main_ind}
Let $F_k=\{s_1, s_2,…,s_l\}$. Then, for any given $k$, there exists a point of time $I_k$ and the constants $c_1,c_2,...,c_l$, where $0<c_1,c_2...,c_l<1$, such that at $I_k$, for each $i\in\{1,...,l\}$ there are $\geq c_in$ agents in $s_i$ a.a.s.
\end{statement}

At this point we want to stress that most of the upcoming lemmas will have a similar statement structure of proving that there exists a constant and a point of time, for which at least a certain fraction of agents is present at some state in this point of time, instead of just saying that $\theta(n)$ agents are reached. This is crucial to ensure the validity of our induction(s). Nevertheless, we will sometimes use the $\theta$-notation for explanatory purposes and better readability.

By Definition \ref{reachable}, in order for the state to be reachable, there must exist a finite configuration sequence, and hence a finite transition sequence, which leads to the configuration having a least one agent in this state. Therefore, Theorem \ref{main_theorem} directly follows from Statement \ref{main_ind} by setting $k$ to the length of the longest such transition sequence among all reachable states in $Q$.

We will prove Statement \ref{main_ind} by induction in $k$.

\vspace{1mm}
\noindent
\textbf{Base step.} For $k=0$ the statement trivially holds, since we can just pick $I_0$ to be the point of time before any interaction occurs, and we have $n$ agents in $s_0$ for any population size $n$. 

\vspace{1mm}
\noindent
\textbf{Inductive hypothesis.}
We assume by Induction that Statement \ref{main_ind} holds for each $k=m$.

\vspace{1mm}
\noindent
\textbf{Inductive step.} Let $k=m+1$ and let $F_k=\{s_1, s_2,…,s_l\}$. Then by ind. assumption there exists a point of time $I_k$, such that there are (possibly more than) $c_1n,c_2n,…,c_ln$ agents in $s_1,s_2,…,s_l$ respectively  ($0<c_1,...,c_l<1$). Let $\{q_1, q_2,…,q_p\} = F_{k+1} \setminus F_k.$ 

Our goal is to show that there exists a point of time $I_{k+1}$, such that there are $\theta(n)$ agents at each of the states $s_1,…,s_l,q_1,…,q_p$. a.a.s. We will show this by a separate induction. The statement of this induction is the following:

\begin{lemma}
    \label{second_ind}
    Let $I_k$ be the point of time of a run of a population protocol \(\mathcal{A}\), such that there are at least $c_1n,c_2n,…,c_ln$ agents in states $s_1,s_2,…,s_l$ respectively ($0<c_1,...,c_l<1$). Let $q_1,q_2,…,q_p$ be the states, such that $\forall a \in\{1,…,l\}, b \in \{1,…,p\}: s_a \neq q_b$ and for each $q_a \in \{q_1,…,q_p\}$ there exist two (possibly same) states $s_i,s_j \in \{s_1,...,s_l\}$, such that there exists at least one of the following transitions: $\delta(s_i,s_j)=(q_a, q)$, $\delta(s_i,s_j)=(q, q_a)$, $q\in Q$. Then there exists a moment of time $I_{k+p}$ and constants $c_1',...,c_l',c_{l+1}',...,c_{l+p}'$, where $0<c_1',...,c_l',c_{l+1}',...,c_{l+p}'<1$, such that at $I_{k+p}$ there are at least $c_1'n,...,c_l'n,c_{l+1}'n,...,c_{l+p}'n$ agents in each of the states $s_1,…,s_l,q_1,...,q_p$ respectively a.a.s. 
\end{lemma}

To put it simple, this lemma states that if at some point of time there are $\theta(n)$ agents in $l$ different states, and $p$ other different states can be reached within one transition from that configuration, with high probability there exists a moment of time when we have $\theta(n)$ agents in all of these $(l+p)$ states. As we see, this directly corresponds to the inductive step of Statement $\ref{main_ind}$.

\vspace{1mm}
\noindent
\textit{Proof of Lemma \ref{second_ind}.} The induction will be in $p$.

\vspace{1mm}
\noindent
\textbf{Base step.} $p=0$. Here we can just set $c_1'=c_1,...,c_l'=c_l$ and $I_{k+p}=I_k$, and use the statement of Lemma \ref{second_ind}.

\vspace{1mm}
\noindent
\textbf{Inductive hypothesis.}
We assume that the statement of Lemma \ref{second_ind} holds for all $p = j$.

\vspace{1mm}
\noindent
\textbf{Inductive step.}
Let $p = j+1$. By ind. hypothesis, at $I_j$ there are $\theta(n)$ agents in the states $s_1,…,s_l,q_1,...,q_p$. Let $q_{p+1}$ be a state, different from any of these states, such that there exists at least one of the following transitions: $\delta(s_i,s_j)=(q_{p+1}, q)$, $\delta(s_i,s_j)=(q, q_{p+1})$, for some $q\in Q$, $s_i,s_j \in \{s_1,…,s_l\}$. Also, let there be at least $c_in$ and $c_jn$ agents in states $s_i$ and $s_j$ respectively at $I_j$, $0<c_i,c_j<1$. 

Let us choose $I_{j+1}$ to be the moment of time, such that there are exactly $\frac{1}{4}(min(c_i,c_j))n$ calls of $ChooseNextPair$ between $I_{j}$ and $I_{j+1}$ for a population size $n$. We will see the reason for such a choice in a moment. Our goal is to prove that at $I_{j+1}$ there are $\theta(n)$ agents at each of the states $s_1,…,s_l,q_1,…,q_p,q_{p+1}$ a.a.s. We will subdivide the proof of this induction step into two separate claims:

\begin{enumerate}
    \item There exists a constant $c_{l+p+1}'$,  $0<c_{l+p+1}'<1$, such that at $I_{j+1}$, at least $c_{l+p+1}'n$ agents will be in $q_{p+1}$ a.a.s. \hfill $(1)$
    \item There exist constants $c_1',...,c_l',c_{l+1}',...,c_{l+p}'$, which lie between $0$ and $1$, such that at least $c_1'n,...,c_l'n$, $c_{l+1}'n,...,c_{l+p}'n$ agents in the states $s_1,…,s_l,q_1,...,q_p$ respectively will not interact between $I_{j}$ and $I_{j+1}$ a.a.s.\hfill $(2)$
\end{enumerate}

To keep the proof modular (and more general), we will prove $(1)$ and $(2)$ as independent lemmas. We will start with $(1)$.

\begin{lemma}
    \label{marking}
    Let \(\mathcal{A}\) be a population protocol with $O(1)$ states that runs on a population \(\mathcal{P}\) of size $n$. Let’s say there is a moment of time $I_1$ of this population protocol's run, such that there are at least $c_in$ and $c_jn$ agents in states $s_i$ and $s_j$ respectively, $0 < c_i,c_j<1$. And let there be a transition on the pair of states $(s_i,s_j)$ in which at least one of the two interacting agents changes its state to $s$. Further, let $I_2$ be the moment of time, such that  $ChooseNextPair$ is called exactly $\frac{1}{4}(min(c_i,c_j))n$ times between $I_1$ and $I_2$. Then, at $I_2$, there are at least $\frac{9}{40}\cdot \frac{5}{256}c_ic_j(min(c_i,c_j))$ agents in state $s$ a.a.s. 
\end{lemma}

\noindent
\textit{Proof of Lemma \ref{marking}.} Let $X$ be the random variable denoting the number of events between $I_1$ and $I_2$ when an agent that was in state $s_i$ at $I_1$ interacts with an agent that was in state $s_j$ at $I_1$.

\vspace{1mm}
\noindent
\textbf{$\alpha$-Marked agents.} Before choosing the agent $a$ in $ChooseNextPair$, we will randomly mark $\frac{1}{4}c_in$ agents in state $s_i$. Similarly, before choosing the agent $b$, we will delete the previously set marks and we will randomly mark $\frac{1}{4}c_jn$ agents in state $s_j$. In the case if $s_i=s_j$, we will mark the agents in a way, such that all the agents marked before choosing $a$ are different from the ones marked before choosing $b$ (still randomly among the possible candidates).  We will call such agents $\alpha-marked$.  All $\alpha$-marks are deleted after the run of $ChooseNextPair$ and the following interaction (in case it happens), and are newly selected in the next call of $ChooseNextPair$. Note that between $I_1$ and $I_2$ we can always do this, since to get less than $\frac{1}{4}c_in+ \frac{1}{4}c_in =\frac{1}{2}c_in$ agents in state $s_i$ more than $\frac{c_in-\frac{1}{2}c_in}{2}=\frac{1}{4}{c_in}$ transitions must occur, and therefore, more than $\frac{1}{4}c_in$ calls of $ChooseNextPair$ (analogously for $s_j$). 

Let $X'$ be a random variable denoting the number of calls of $ChooseNextPair$ between $I_1$ and $I_2$, such that:

\begin{enumerate}
    \item One of the two chosen agents was in state $s_i$ at $I_1$, and the other one was in state $s_j$ at $I_1$;
    \item Each of these agents was $\alpha$-marked before the call.
\end{enumerate}

Note that $X\geq X'$, since if agents $a$ and $b$ are distinct, $\alpha$-marked agents form a subset of all agents in states $s_i$ and $s_j$, and otherwise, this event is not counted towards the both random variables.  Since the number of agents in both states that are $\alpha$-marked is the same before every call, the probability of each call of $ChooseNextPair$ between $I_1$ and $I_2$ to be counted towards $X'$ is $\frac{\frac{1}{4}c_in}{n}\cdot  \frac{\frac{1}{4}c_jn}{n}=\frac{1}{16}c_ic_j$ both if $s_i \neq s_j$, and if $s_i=s_j$. Therefore, $X'\sim Bin(\frac{1}{4}(min(c_i,c_j))n, \frac{1}{16}c_ic_j)$. 

\vspace{1mm}
We now want to bound the probabilities that $X'$ strongly deviated from its expected value. For this, we will make use of Chebyshev’s inequality, for which we have reasonably chosen the constants, in order for the proof to work. As a remark, we will note that particularly Chebyshev's inequality restricts us to relatively small margins for these constants in this case. Some other inequalities, such as the Chernoff bound or Cantelli's inequality (one-sided version of Chebyshev's inequality), allow for larger margins and possibly somewhat stronger results, but we decided to stick to Chebyshev's bounds here and in the following proofs in order to slightly simplify the calculations. Note that the main claim can also be proved by choosing the constants different from those chosen in the below inequalities. We have:
\begin{equation*}
    \hspace*{-0.8 cm}Pr[|X'-\frac{1}{64}c_ic_j(min(c_i,c_j))n|\geq \frac{1}{256}c_ic_j(min(c_i,c_j))n] 
\end{equation*}
\begin{equation*}
    \leq\frac{\frac{1}{4}(min(c_i,c_j))n\frac{1}{16}c_ic_j(1-\frac{1}{16}c_ic_j)}{(\frac{1}{256}c_ic_j(min(c_i,c_j))n)^2}=\frac{1024(1-\frac{1}{16}c_ic_j)}{(min(c_i,c_j))c_ic_j}\cdot\frac{1}{n}. 
\end{equation*}

Hence,

\begin{equation}
    \label{lower_x}
    Pr[X'<\frac{3}{256}c_ic_j(min(c_i,c_j))n] \leq \frac{1024(1-\frac{1}{16}c_ic_j)}{(min(c_i,c_j))c_ic_j}\cdot\frac{1}{n}.
\end{equation}

At the same time:
\begin{equation}
    \label{upper_x}
    Pr[X'\geq\frac{5}{256}c_ic_j(min(c_i,c_j))n] \leq \frac{1024(1-\frac{1}{16}c_ic_j)}{(min(c_i,c_j))c_ic_j}\cdot \frac{1}{n}.
\end{equation}

Let $Y$ be a random variable denoting the number of agents that reached state $s$ as a result of an interaction between an agent in state $s_i$ and an agent in state $s_j$, and then interacted at least once again - all between $I_1$ and $I_2$. In the following, we aim to give the high-probability upper bound for $Y$, using the same idea of marking the agents as we did before.

\vspace{1mm}
\noindent
\textbf{$\beta$-Marked agents.} Before each run of $ChooseNextPair$, we denote $\#(s)$ to be the number of agents in state $s$, that have reached this state a result of an interaction between agents in states $s_i$ and $s_j$ between $I_1$ and $I_2$. We mark $cn$ agents before each execution of $ChooseNextPair$, where $c=\frac{5}{256}c_ic_j(min(c_i,c_j))$, as follows: 

\begin{itemize}
    \item if $\#(s)\leq cn$, we will randomly mark $(cn-\#(s))$ agents and all the agents that are in state $s$ (and have reached this state as a result of an interaction between an agent in state $s_i$ and $s_j$ between $I_1$ and $I_2$);
    \item otherwise, if $\#(s) > cn$, we will randomly mark $cn$ agents from the whole population.
\end{itemize}

We will call such agents $\beta-marked$.  All $\beta$-marks are deleted after the execution of each $ChooseNextPair$ procedure, and are newly selected before the next one. 


Let $Y'$ be a random variable denoting the number agents chosen by $ChooseNextPair$, that were $\beta$-marked, between $I_1$ and $I_2$. In the case when both $a$ and $b$ agents are $\beta$-marked, each of them is counted towards $Y'$. Since the probability of each chosen agent to be marked is $c$, and each $ChooseNextPair$ drafts two agents, $Y'\sim Bin(\frac{1}{2}(min(c_i,c_j))n, c)$. Note that $c$ is independent of the actual value of $X'$, which ensures the independence of trials.

By Chebyshev's inequality, 
\begin{equation*}
    Pr[|Y'-\frac{1}{2}c(min(c_i,c_j))n|\geq \frac{1}{4}c(min(c_i,c_j))n]\leq \frac{\frac{1}{2}(min(c_i,c_j))nc(1-c)}{(\frac{1}{4}c(min(c_i,c_j))n)^2}= \frac{8(1-c)}{c(min(c_i,c_j))} \cdot \frac{1}{n}.
\end{equation*}

Hence,
\begin{equation}
    \label{upper_y}
    Pr[Y'>\frac{3}{4}c(min(c_i,c_j))n]\leq \frac{8(1-c)}{c(min(c_i,c_j))} \cdot \frac{1}{n}.        
\end{equation}

Note that $X-Y$ is the lower bound for the number of agents that are in state $s$ at $I_2$. For $c=\frac{5}{256}c_ic_j(min(c_i,c_j))$, by using Bayes' theorem, we have: 

\begin{equation*}
    \hspace*{-5.3 cm} Pr[X-Y\geq c(\frac{3}{5}-\frac{3}{4}(min(c_i,c_j)))n] 
\end{equation*}
\begin{equation*}
    \hspace*{-1.2 cm}=\frac{Pr[X-Y\geq c(\frac{3}{5}-\frac{3}{4}(min(c_i,c_j)))n|X'<cn] \cdot Pr[X'<cn]}{Pr[X'<cn|X-Y\geq c(\frac{3}{5}-\frac{3}{4}(min(c_i,c_j)))n]}
\end{equation*}
\begin{equation*}
    \geq Pr[X-Y\geq c(\frac{3}{5}-\frac{3}{4}(min(c_i,c_j)))n|X'<cn] \cdot Pr[X'<cn] = (*)
\end{equation*}

Our goal is to show the high probability of having a non-zero fraction of agents in state $s$ at $I_2$, namely, at least  $c(\frac{3}{5}-\frac{3}{4}(min(c_i,c_j)))n$ agents. Note that $c(\frac{3}{5}-\frac{3}{4}(min(c_i,c_j)))\geq$

\vspace{1mm}
\noindent
$c(\frac{3}{5}-\frac{3}{4}\cdot \frac{1}{2})=\frac{9}{40}\cdot c>0$, since $c_i+c_j\leq1$ and $c_i,c_j>0$.

\vspace{1mm}
Further note that under $X'<cn$, $\#(s)$ is always less or equal than $cn$ between $I_1$ and $I_2$, which means that $Y'\geq Y$ in this case. Besides, we always have $X'\leq X$, which gives us:

\begin{equation*}
     \hspace*{-1.5 cm} (*) \geq Pr[X'-Y'\geq c(\frac{3}{5}-\frac{3}{4}(min(c_i,c_j)))n|X'<cn] \cdot Pr[X'<cn]
\end{equation*}
\begin{equation*}
    \hspace*{-1.6 cm}\geq Pr[X'-Y'\geq c(\frac{3}{5}-\frac{3}{4}(min(c_i,c_j)))n \wedge X'<cn] \cdot Pr[X'<cn]
\end{equation*}
\begin{equation*}
    \hspace*{-0.2 cm}\geq Pr[X'\geq \frac{3}{5}cn  \wedge Y'\leq \frac{3}{4}c(min(c_i,c_j))n \wedge X'<cn] \cdot Pr[X'<cn] = (**). 
\end{equation*}

Further, by Fréchet \cite{frechet1935generalisation} inequalities we have:
\begin{equation*}
     \hspace*{-1.9 cm}(**)\geq (Pr[X'\geq \frac{3}{5}cn]+Pr[Y'\leq \frac{3}{4}c(min(c_i,c_j))n] + Pr[ X'<cn]- 2)\cdot Pr[X'<cn]
\end{equation*}
\begin{equation*}
    =(1 - Pr[X'< \frac{3}{5}cn]+1 - Pr[Y'> \frac{3}{4}c(min(c_i,c_j))n]+1-Pr[X'\geq cn] -2)\cdot (1 -Pr[X'\geq cn])
\end{equation*}
\begin{equation*}
    \hspace*{-3.8 cm}\geq (1 -  \frac{1024(1-\frac{1}{16}c_ic_j)}{(min(c_i,c_j))c_ic_j}\cdot\frac{1}{n}- \frac{8(1-c)}{c(min(c_i,c_j))} \cdot \frac{1}{n}-\frac{1024(1-\frac{1}{16}c_ic_j)}{(min(c_i,c_j))c_ic_j}\cdot \frac{1}{n})
\end{equation*}
\begin{equation}
    \label{ineq_frechet}
    \hspace*{3.6 cm}\times (1 - \frac{1024(1-\frac{1}{16}c_ic_j)}{(min(c_i,c_j))c_ic_j}\cdot \frac{1}{n}).
\end{equation}

The last inequality follows from Inequalities \ref{lower_x}, \ref{upper_x} and \ref{upper_y}. 

To conclude, we have shown that there exists a moment of time $I_2$ and three constants
\begin{itemize}
    \item[] $const_1=c(\frac{3}{5}-\frac{3}{4}(min(c_i,c_j)))$,
    \item[] $const_2=\frac{1024(1-\frac{1}{16}c_ic_j)}{(min(c_i,c_j))c_ic_j}+\frac{8(1-c)}{c(min(c_i,c_j))} +\frac{1024(1-\frac{1}{16}c_ic_j)}{(min(c_i,c_j))c_ic_j}$, and
    \item[] $const_3=\frac{1024(1-\frac{1}{16}c_ic_j)}{(min(c_i,c_j))c_ic_j}$,
\end{itemize}

where $c=\frac{5}{256}c_ic_j(min(c_i,c_j))$, such that $0<const_1, const_2, const_3<1$, and:

\begin{equation*}
    Pr[X-Y\geq const_1n]\geq (1-const_2 \cdot \frac{1}{n})(1-const_3 \cdot \frac{1}{n})\rightarrow 1, n \rightarrow \infty.
\end{equation*}

From here, it follows that there are at least $\frac{9}{40}\cdot \frac{5}{256}c_ic_j(min(c_i,c_j))$ agents in state $s$ a.a.s. \hfill{$\square$}

\vspace{1mm}
Before showing Claim $(2)$, we will prove the following lemma.

\begin{lemma}
    \label{untouched}
    Let \(\mathcal{A}\) be a population protocol with $n$ agents and $O(1)$ states, and let $I_1$ be the moment of time at which there are at least $c_in$ agents in state $s_i$, $0<c_i<1$. We call an agent $\textbf{untouched}$ if he has not interacted since $I_1$. Let $I_2$ be the moment of time when $tn$ calls of $ChooseNextPair$ happened for some $t$, $0<t<1$. Then, $Pr[U_i< (1-3t)c_in]\rightarrow 0$, $n\rightarrow \infty$, where $U_i$ is the number of untouched agents in state $i$ at $I_2$.
\end{lemma}

\noindent
\textit{Proof of Lemma \ref{untouched}.} Let $X_i$ be a random variable denoting the number agents chosen by $ChooseNextPair$ between $I_1$ and $I_2$, that were in $s_i$ at $I_1$. Since each $ChooseNextPair$ procedure chooses two agents, the described events are independent with success probability $c_i$, and $X_i\sim Bin(2tn, c_i)$.  By Chebyshev’s inequality we have:

\begin{equation*}
    Pr[|X_i-2tc_in|\geq tc_in]\leq \frac{2tnc_i(1-c_i)}{(tc_in)^2}=\frac{2(1-c_i)}{tc_i}\cdot \frac{1}{n}
\end{equation*}
\begin{equation}
    \label{x_upper_2}
    \hspace*{-3.3 cm}\Longrightarrow Pr[X_i> 3tc_in]\leq \frac{2(1-c_i)}{tc_i}\cdot \frac{1}{n}
\end{equation}

Let $Y_i$ be the count of agents that were at $s_i$ at $I_1$ and interacted at least once between $I_1$ and $I_2$. Note that $X_i\geq Y_i$ $(***)$,  since every interaction of an agent that increases the value of $Y_i$ is also counted towards $X_i$. Further, for any random variables $X,Y,Z$:
\begin{equation*}
    \hspace*{-5 cm}Pr[Y>Z] = \sum_{y\in Y, z\in Z}(y>z)Pr[Y=y,Z=z]
\end{equation*}
\begin{equation*}
    \hspace*{0.8 cm}\leq \sum_{x\in X,y\in Y,x\in Z}((y>x) \vee (x > z)) Pr[X=x,Y=y,Z=z]
\end{equation*}
\begin{equation*}
    \hspace*{-4.4 cm}=Pr[(Y>X) \vee (X>Z)]
\end{equation*}
\begin{equation}
\label{xyz}
    \hspace*{-4.2 cm}\leq Pr[Y>X]+Pr[X>Z].
\end{equation}

The first inequality holds, since $\forall x,y,z\in \mathbb{R}$ : $y>z\Rightarrow(y>x)\vee(x>z)$. For the last one we used the union bound. From the Inequalities \ref{x_upper_2}, $(***)$ and \ref{xyz} now follows:

\begin{equation*}
    Pr[Y_i >3tc_in]\leq Pr[Y_i>X_i]+Pr[X_i>3tc_in]\leq 0 + \frac{2(1-c_i)}{tc_i}\cdot \frac{1}{n}
\end{equation*}
Note that $U_i = c_in - Y_i$. Hence, $Pr[U_i<(1-3t)c_in]\leq\frac{2(1-c_i)}{tc_i}\cdot \frac{1}{n} \rightarrow 0$, $n\rightarrow \infty$. \hfill{$\square$}

\vspace{1mm}
Since $(1)$ is a direct corollary of Lemma 1, it remains to show $(2)$ and combine the both claims for our induction step.

As in Lemma \ref{untouched}, we call an agent $untouched$ if he has not interacted since $I_1$. To generalize $(2)$, we will show the following statement (in terms of previous definitions of our population protocol \(\mathcal{A}\) , and states $c_i$ and $c_j$):

\begin{statement}
    \label{statement_untouched}
    There exist constants $c_1',...,c_l'$, which lie between $0$ and $1$, such that at least $c_1'n,...,c_l'n$ agents in the states $s_1,…,s_l$ respectively will not interact between $I_1$ and $I_2$ a.a.s., if at $I_1$ there were at least $c_1n,…,c_ln$ agents in these states respectively. (As previously, there are $\frac{1}{4}(min(c_i,c_j))n$ calls of $ChooseNextPair$ between $I_1$ and $I_2$).
\end{statement}

\textit{Proof of Statement \ref{statement_untouched}.} It is enough to show, that there exists a constant $const$, $0<const<1$, such that for $n\rightarrow \infty$, the probability that for at least one index $i$,  $U_i<const \cdot n$, goes to 0 a.a.s., where $U_i$ is the number of untouched agents in state $i$ at $I_2$. Or formally:

\begin{equation*}
    \exists const: Pr[\bigcup_{i=1}^l (U_i< const \cdot n)]\rightarrow 0, n\rightarrow \infty.
\end{equation*}

For each index $i$ applying Lemma \ref{untouched} for $t=\frac{1}{4}(min(c_i,c_j))$, gives:

\begin{equation*}
    Pr[U_i<(1-3\cdot \frac{1}{4}(min(c_i,c_j)))c_in]\leq\frac{2(1-c_i)}{\frac{1}{4}(min(c_i,c_j))c_i}\cdot \frac{1}{n}.
\end{equation*}

By taking the union bound for all states we have: 

\begin{equation*}
    \hspace*{-0.3 cm}Pr[\bigcup_{i=1}^k (U_i<(1-3\cdot \frac{1}{4}(min(c_i,c_j)))c_in]
\end{equation*}
\begin{equation*}
    \leq \sum_{i=1}^n Pr[U_i<(1-3\cdot \frac{1}{4}(min(c_i,c_j)))c_in]
\end{equation*}
\begin{equation}
    \label{ineq_union}
    \hspace*{-2.6 cm}\leq \sum_{i=1}^k \frac{2(1-c_i)}{\frac{1}{4}(min(c_i,c_j))c_i}\cdot \frac{1}{n}
\end{equation}

Setting $const= (1-3\cdot \frac{1}{4}(min(c_i,c_j)))c_i$, and $const_2=\sum_{i=1}^k \frac{2(1-c_i)}{\frac{1}{4}(min(c_i,c_j))c_i}$ ($const_2$ is a constant only depending on constants $c_i$ and $c_j$, $0<const_2<1$) gives:

\begin{equation*}
    Pr[\bigcup_{i=1}^k (U_i<const \cdot c_in)]\leq const_2\cdot \frac{1}{n}\rightarrow 0, n\rightarrow \infty,
\end{equation*}
which concludes the proof.\hfill{$\square$}

\vspace{1mm}
Getting back to the Induction Step of Lemma \ref{second_ind}, we now need to combine the statements $(1)$ and $(2)$ to infer its proof. The statements individually follow from Lemma \ref{marking} and Statement \ref{statement_untouched} respectively, for $I_1=I_j$ and $I_2=I_{j+1}$. 
Note that we have used the same number of interactions between $I_1$ and $I_2$ in the both proofs, which means that we can take our final union bound in order to show that there is a non-zero fraction of agents in each of the $s_1,…,s_l,q_1,…,q_p,q_{p+1}$ a.a.s. at $I_{j+1}$. By using Inequalities \ref{ineq_frechet} and \ref{ineq_union}:

\begin{equation*}
    \hspace*{-2.5 cm}Pr[\neg(1) \vee \neg(2)]\leq Pr[\neg (1)]+Pr[\neg (2)]=1 -Pr[(1)]+Pr[\neg (2)]
\end{equation*}
\begin{equation*}
    \hspace*{1.9 cm}\leq 1 - (1-const_2 \cdot \frac{1}{n})(1-const_3 \cdot \frac{1}{n}) + const_4\cdot \frac{1}{n} \text{, where}
\end{equation*}

$const_2$ and $const_3$ correspond to the same-named constants from Inequality \ref{marking}, and $const_4= \sum_{i=1}^k \frac{2(1-c_i)}{\frac{1}{4}(min(c_i,c_j))c_i}$. Hence, 

\begin{equation*}
    Pr[(1) \wedge (2)]\geq (1-const_2 \cdot \frac{1}{n})(1-const_3 \cdot \frac{1}{n}) - const_4\cdot \frac{1}{n}\rightarrow 1, n\rightarrow \infty.
\end{equation*}

Now we have shown the inductive step and have all the prerequisites needed for the $(k+2)$nd step.

Finally, by the Principle of Mathematical Induction the statement of Lemma \ref{second_ind} holds for all $p$. This means that we can choose $I_{k+p}$ to be the moment of time, when there had been exactly $\sum_{m=1}^{p} \frac{1}{4}(min(c_i^{m},c_j^{m}))n$ calls of $ChooseNextPair$ starting from $I_k$, where $c_i^m$ and $c_j^m$ denote the fraction of agents in two states among $s_1,…,s_l$, from which the state $q_{m}$ can be achieved, at the moment of time, when there had been exactly $\sum_{q=1}^{m-1} \frac{1}{4}(min(c_i^{q},c_j^{q}))n$ calls of $ChooseNextPair$ starting from $I_k$.\hfill{$\square$}
\vspace{1mm}

Getting back to the proof of Theorem \ref{main_theorem}, we have now shown its Inductive Step too. By the Principle of Mathematical Induction, for any given $k$, there exist a point of time $I_k$, such that at $I_k$ there are $\theta(n)$ agents at each state in $F_k$ a.a.s. \hfill{$\blacksquare$}  

\vspace{1mm}
To conclude, we have proven that there exists a point of time $I$, in which all reachable states possess a non-zero fraction of population with high probability. We separate out a stronger statement in form of the following immediate corollary.

\begin{corollary}
    The point of time $I$ from Theorem \ref{main_theorem} occurs within $\theta(n)$ interactions, starting from the beginning of protocol's run a.a.s.
\end{corollary}

\textit{Proof.} To be precise, the statement of Theorem \ref{main_theorem} subsequently holds when we choose $I$ to be moment of time, when there had been exactly $\sum_{h=0}^{h_{max}-1}\sum_{m=1}^{p_h} \frac{1}{4}(min(c_i^{m},c_j^{m}))n:=T_{calls}$ calls of $ChooseNextPair$ from the start of protocol’s run, where $p_h=|F_{h+1}\setminus F_h|$, and $c_i^m$ and $c_j^m$ are defined, as previously, in terms of the corresponding sets $F_{h+1}\setminus F_h$. Although $c_i^{m}$ and $c_j^{m}$ are defined recursively, they only depend on the initial fractions of agents in the states, and the number of states, but are constant w.r.t. $n$. This yields $T_{calls}=\theta(n)$.  

The probability of $ChooseNextPair$ not resulting into a respective interaction afterwards is $\frac{1}{n}$, since it happens exactly when the same agent is chosen twice. Hence, $E[T_{interactions}]=(1-\frac{1}{n})E[T_{calls}]=\theta(n) - \frac{1}{n}\cdot \theta(n)=\theta(n)$, where $T_{interactions}$ is the expected number of interactions before we have $\theta(n)$ agents in each reachable state.\hfill{$\square$} 
\vspace{1mm}

We can now apply our main result to the question of confident leader election.

\begin{corollary}
    There does not exits a probabilistic population protocol that solves Confident leader election problem with high probability.
\end{corollary}

\textit{Proof.} Due to Definition \ref{conf_leader}, a prerequisite of having a confident leader in a population protocol, is to have a set of reachable states (containing at least one state) of sub-type $CL$, that are only visited by one agent with high probability. However, Theorem \ref{main_theorem} claims that each reachable state will eventually contain a non-zero fraction of population at the same time, which contradicts the above prerequisite.\hfill{$\square$} 
\chapter{Further Experiments}\label{chapter:experiments}

A specific lower bound for the number of agents that will reach a certain state due to Theorem \ref{main_theorem} highly depends on the layer $k$, i.e. $F_{k}\setminus F_{k-1}$ in terms of notations used in Statement $\ref{main_ind}$, in which this state lies with respect to the protocol. Lemma \ref{marking} shows, that the fraction-constant gets at least squared for each next layer, which yields a doubly-exponential expression for the high-probability lower bound for the amount of agents in a state of layer $k$ in the best case. 

Although it is still enough to show our main result, in practice, in order to guarantee even at least one agent in a state of layer $3$ with high probability using this bound, the population size will have to be at least 

\begin{equation*}
    \frac{1}{(\frac{9}{40}\cdot \frac{5}{256}c_ic_j(min(c_i,c_j)))^3}\geq \frac{1}{(\frac{9}{40}\cdot \frac{5}{256}\cdot \frac{1}{2}\cdot \frac{1}{2}\cdot \frac{1}{2})^3} > 6\cdot 10^9.
\end{equation*}
\vspace{1mm}

\begin{figure}[htp]
    \centering
    \includegraphics[width=250pt]{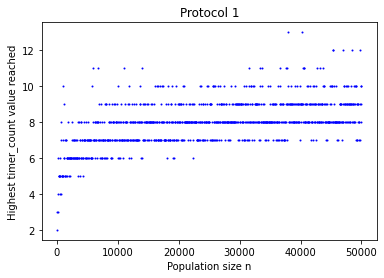}
    \caption{Simulation results: highest value of timer\_count}
    \label{fig:Sim_2}
\end{figure}

A natural question here is how close our expression is to being optimal. While it is hard to give a comprehensive, theory-driven answer to this question directly, we decided to make use of our implementation of Protocol $1$ for leader election from Chapter 3 to understand how large the gap to the optimal expression can potentially be. We ran Protocol $1$ once for each value of $(10 + 50n)$ between $10$ and $50000$ for $n = 0,1,2,...$ \cite{gitlab}. This time we waited until the single leader is elected and did not terminate the protocol early. We kept track of the highest value of timer\_count value among all agents throughout the whole protocol run. The results are depicted in Figure \ref{fig:Sim_2}.

Note, that timer\_count equal to $k$ corresponds to a state that lies at least in the layer $k$. Even though the results are quite noisy, it can be seen that the relation between the highest $k$ reached and $n$ is very different from the one we would get by just applying the lower bound estimation described above. Moreover, one could recognize a single-logarithmic-like relation in the contrary to the doubly-logarithmic relation that can be inferred from Lemma \ref{marking}. 

This leads us to an open question of whether it can be guaranteed with high probability that the state of layer $k$ after $\theta(n)$ time will have a fraction of agents of at least $c^k$ for a constant $c$. At the same time, we do not exclude the possibility (but rather find it quite plausible) of existence of a population protocol, that is able to find a confident leader, in the case if the relation between $k$ and $n$ is even doubly logarithmic. One of the ideas could be to consider a protocol, that is based on Protocol $1$ with the difference that the counter agents also track the number of consecutive interactions with leaders. In case if a leader and a counter spot a mismatch in their interactions-with-the-opposite-type counts, they reset (here by opposite types we mean leaders and counters).
\chapter{Conclusion}\label{chapter:conclusion}

\section{Achieved Results}
To conclude, in this thesis we have proven the negative answer to an open question of the possibility of solving the Confident leader election problem with probabilistic population protocols. As a more general result, we have shown that for a population protocol with a constant amount of states there exists a moment of time that lies within $\theta(n)$ interactions from the start of protocol's run, such that each reachable state is occupied by $\theta(n)$ agents a.a.s. Some of the lemmas, that we presented while proving our main result may be adapted in the future work on studying the population protocols, as they capture one of the aspects of evolution of the number of agents in particular states.

\section{Open Questions and Future Work}

Note, that we do not give:

\begin{itemize}
    \item  a specific upper bound on the number of interactions before $I$ (from the statement of Theorem \ref{main_theorem}), and
    \item  a specific lower bound on the number of agents in a reachable state of layer $k$ at the point of time $I$.
\end{itemize}

These parameters can be closer investigated in future work. 

As we argued in Chapter 5, our worst-case lower bound is very likely to be far from the optimal one. This means, that for a large number of combinations of the layer $k$ in which the state detecting a confident leader lies, and the population size $n$, where $k$ is relatively large, and $n$ - relatively small, our proof does not yet guarantee that even at least two agents reach the confident leader state with high probability. This was also demonstrated with an example in Chapter 5. Hence, we pose the two main open question as follows:

\begin{enumerate}
    \item Does there exist a state-efficient population protocol that is capable with high probability of solving the Confident leader election problem for small $n$? If yes, what is the optimal achievable relation between $k$ and $n$?
    \item What is the best-possible high probability lower bound on the number of agents that will visit a state of layer $k$ assuming the population of size $n$?
\end{enumerate}


Finally, we have not addressed the question of how the statement of the proof, as well as the respective high probability bounds, change in case if the initial configuration contains agents in more than one state, or, say, is $\alpha$-dense (meaning that each state present in the configuration is occupied by at least $\alpha n$ agents).


\appendix{}


\microtypesetup{protrusion=false}
\microtypesetup{protrusion=true}
\printglossaries
\printbibliography{}

\end{document}